# Exciting Andreev pairs in a superconducting atomic contact

L. Bretheau,[1*] Ç. Ö. Girit,[1*] H. Pothier,[1] D. Esteve[1] and C. Urbina[1]

[1]Quantronics Group, Service de Physique de l'État Condensé (CNRS, URA 2464), IRAMIS, CEA-Saclay, 91191 Gif-sur-Yvette, France

[*]These authors contributed equally to this work.

**The Josephson effect describes the flow of supercurrent in a weak link, such as a tunnel junction, nanowire, or molecule, between two superconductors[1]. It is the basis for a variety of circuits and devices, with applications ranging from medicine[2] to quantum information[3]. Currently, experiments using Josephson circuits that behave like artificial atoms[4] are revolutionizing the way we probe and exploit the laws of quantum physics[5,6]. Microscopically, the supercurrent is carried by Andreev pair states, which are localized at the weak link. These states come in doublets and have energies inside the superconducting gap[7-10]. Existing Josephson circuits are based on properties of just the ground state of each doublet and so far the excited states have not been directly detected. Here we establish their existence through spectroscopic measurements of**

**superconducting atomic contacts. The spectra, which depend on the atomic configuration and on the phase difference between the superconductors, are in complete agreement with theory. Andreev doublets could be exploited to encode information in novel types of superconducting qubits [11-13].**

A bulk, isolated Bardeen-Cooper-Schrieffer (BCS) superconductor can be described by a spectrum having a gap around the Fermi energy of $2\Delta$, which is the minimum energy necessary to excite a pair[14]. In the presence of a short weak link, the superconducting phase $\delta$ can be easily twisted, leading to a local modification of the spectrum and the creation of new states inside the gap. These Andreev Bound States have energies $\mp E_A$, with $E_A$ given by

$$E_A = \Delta\sqrt{1-\tau\sin^2(\delta/2)} \qquad (1)$$

for a weak link which has a single conduction channel of transmission probability $\tau$ (Figure 1a). Since $E_A \leq \Delta$, these bound states cannot propagate into the bulk superconductor and are localized at the weak link on a distance of order $\xi$, the superconducting coherence length. The ground Andreev pair state $|-\rangle$ has energy $-E_A$ and the lowest possible pair excitation of the system, requiring an energy $2E_A$, is a transition to the excited Andreev pair state $|+\rangle$ at $+E_A$. The phase dependence of $\mp E_A$ gives rise to opposite supercurrents for the two states, $\mp(2\pi/\phi_0)(\partial E_A/\partial \delta)$, with $\phi_0=h/2e$ the flux quantum.

Current Josephson circuits are primarily based on tunnel Josephson junctions, which have many conduction channels with small transmissions ($\tau \ll 1$). In this limit, the ground state energy $-E_A$ in each channel is proportional to $-\cos \delta$. Summing over all channels one recovers the standard Josephson coupling energy $-E_J \cos \delta$ and the sinusoidal current-phase relation predicted by Josephson[1]. For channels of arbitrary transmissions, the ground state $|-\rangle$ has been probed through measurements of the current-phase relation in superconducting atomic contacts[15]. Excitations created by the addition or removal of an electron from the state $|-\rangle$ have been observed in superconducting atomic contacts[16] and quantum dots connected to superconductors[17,18]. The continuum of Andreev states that form in SNS structures has also been probed[19,20]. Thermal occupation of the excited states was invoked to explain the temperature dependence of the supercurrent[10]. However, the excited Andreev pair state $|+\rangle$ has not been directly detected. Here we present spectroscopic evidence of excited Andreev pair states in superconducting atomic contacts, a simple system that allows direct quantitative comparison with theoretical predictions.

The principle of our experiment is described in Figure 1b. An atomic contact obtained using a microfabricated, mechanically-controllable break junction[21] is placed in parallel with a tunnel Josephson junction to form a SQUID. A second tunnel junction, the "spectrometer", is used as an on-chip broadband microwave source and detector[22-24]. It is coupled to the SQUID through an on-chip capacitor ($\sim 30$ pF). The superconducting material for the junctions and atomic contact is aluminum ($\Delta \simeq 180\,\mu\mathrm{eV}$) (see Methods for fabrication details). A micrograph of the sample is shown in Figure 1d. Both the spectrometer and the SQUID can be voltage-biased separately through on-chip LC filters (see Supplementary Figures 1&2). The transmissions of the conduction channels of the atomic contact are determined by fitting the current-voltage characteristic of the SQUID with the theory of multiple Andreev reflections[25] (see Supplementary Figure 3). The SQUID geometry also allows phase biasing the atomic contact by applying a magnetic flux $\phi$ through the loop. Since the sum of the Josephson inductance of the SQUID tunnel junction (~310 pH) and the inductance of the SQUID loop (~20 pH) is much smaller than the typical atomic contact inductance (~3 nH), the phase difference across the atomic contact is $\delta \approx \varphi = 2\pi\phi/\phi_0$.

When biased at a voltage $V_J$ the spectrometer undergoes Josephson oscillations and acts as a microwave current source at frequency $\nu_J = 2eV_J/h$. Microwave photons emitted by the spectrometer are absorbed by the environment which subsequently relaxes. The dissipated power $P$ requires a dc current $I_J$ to be supplied by the biasing circuit to satisfy power conservation $P = I_J V_J$. Microscopically, this dc current is a result of inelastic Cooper-pair tunneling: each time a photon is absorbed, a Cooper pair tunnels across the spectrometer insulating barrier[26,27], as in Figure 1c. In the sub-gap current-voltage $I_J(V_J)$ characteristic of the spectrometer junction, a transition of energy $E$ is revealed as a dc current peak at $2eV_J = E$ with height $I_J = 2e\Gamma(E)$, where $\Gamma(E)$ is the photon absorption rate. Classically, this rate is related to the real part of the impedance seen by the spectrometer. The on-chip coupling capacitor and LC filtering are designed to keep the absorption rate due to the external environment low. Transitions such as the Andreev excitation $\left(|-\rangle \to |+\rangle\right)$ at the energy $2eV_J = 2E_A(\delta,\tau)$ can be distinguished by their dependence on both the flux and the contact configuration.

Figure 2 shows the current-voltage characteristic $I_J(V_J)$ of the spectrometer for atomic contact AC2 (see below) at two values of the reduced flux $\varphi$. Several current peaks are visible below the voltage $\Delta/e \simeq 180\,\mu$V, which corresponds to the maximum excitation energy of interest for the Andreev transition, $2eV_J = 2\Delta$. Parts of the spectra change by as much as 200 pA as a function of reduced flux $\varphi$, revealing excitation of modes associated with the SQUID. Specifically, when going from $\varphi=1.15\pi$ (black line) to $\varphi=\pi$ (red line), a prominent peak develops at a voltage bias $V_J=20$ µV and a peak at ~90 µV broadens. Peaks which do not depend on the flux bias or the contact configuration, for example around $V_J=150$ µV, are interpreted as resonances in the external electromagnetic environment and form a background which is subtracted from the IVs (see section 2.2 of Supplementary information). In Figure 2b there are no data in the two grey regions ($V_J \approx 50\,\mu\text{V}$ and $V_J \leq 9\,\mu\text{V}$) because the spectrometer voltage biasing is not stable (see Methods). The measured current decreases about an order of magnitude as the bias voltage is increased and passes through the zone of instability at $V_J \approx 50\,\mu\text{V}$.

Spectra measured for the three different atomic contacts AC1, AC2 and AC3 are shown in Figures 3a-c. In each spectrum, the current $I_J$ through the spectrometer junction is plotted with the common color scale of Figure 3c. The vertical axes give the energy of photons emitted by the spectrometer in units of the bias voltage, $h\nu_J=2eV_J$. The corresponding frequencies range from 0 to 85 GHz. The horizontal axes give the applied reduced flux $\varphi \approx \delta$. There are no data in the grey regions where biasing is unstable. The contrast becomes fainter as the energy increases, except for a narrow band around 1.8Δ. The most remarkable features are the V-shaped transitions which fan out from $\varphi=\pi$ towards higher energies. AC3, which is a many-atom contact with about 20 conduction channels (Figure 3c), has a multitude of well resolved V-shaped transitions. These transitions, which depend sensitively on the channel transmissions $\tau_i$ as well as $\varphi$, are the Andreev transitions. To confirm this, we plot with red lines in Figures 4b (AC1) and 4d (AC2) the expected positions $2E_{A1}$ and $2E_{A2}$ of the Andreev transitions using Eq. 1 for the two highest transmission channels in each contact: AC1 (transmissions 0.942, 0.26) and AC2 (transmissions 0.985, 0.37) (see Supplementary Figure 3). The lines match the observed transitions.

In addition to the Andreev transitions, there is a much brighter spectroscopic line ($I_J >1$ nA, color scale red) common to all contacts which is located at $0.51\Delta$ and hardly varies with flux for AC1 and AC2 but dips to $0.4\Delta$ at $\varphi \sim \pi$ for AC3. It corresponds to the large peak at $V_J=45$ µV in Figure 2, whose upper half falls in the region of instability. We identify it as the excitation of the plasma mode of the SQUID. This oscillator mode formed by the SQUID Josephson inductance $L_S(\varphi)$ and its parallel capacitance $C_S$ resonates at frequency $\nu_p = \left(2\pi\sqrt{L_S C_S}\right)^{-1} \approx 22\,\text{GHz}$ ($0.51\Delta/h$). The capacitance $C_S \simeq 280\,\text{fF}$ is the sum of the SQUID and spectrometer capacitances. $L_S(\varphi)$ results from the parallel combination of three inductive elements: the atomic contact, the SQUID Josephson junction, and an on-chip inductor on the biasing line (see Supplementary Figure 1). The flux dependence of $L_S(\varphi)$ is negligible for the asymmetric SQUIDs (cases AC1, AC2) but results in a $0.1\Delta$ amplitude modulation (4 GHz) of the plasma frequency for the large atomic-contact SQUID (case AC3). The energy $h\nu_p$ associated with the plasma frequency $\nu_p$ for AC1 and AC2 is plotted in Figures 4b and 4d, respectively, as blue lines, and concurs with the experimental data. The abrupt decrease in spectrometer signal above the plasma frequency (Figure 2b) is due to the shunting of emitted microwaves by the capacitance $C_S$.

The combination of the Andreev and plasma degrees of freedom leads to a double ladder energy diagram as shown in Figure 4a. The states are labeled by $|\sigma,n\rangle$, where $\sigma=\pm$ accounts for the Andreev pair state and $n$ is the plasmon number. The data are well explained by considering transitions only from the initial state $|-,0\rangle$. The Andreev transition $(|-,0\rangle \to |+,0\rangle)$ at $2eV_J = 2E_A$ is indicated by the red arrow and the plasma transition $(|-,0\rangle \to |-,1\rangle)$ at $2eV_J = h\nu_p$ by the blue arrow.

In the spectrum of each contact, there is another resonance near $1.02\Delta$ which is similar in shape to the plasma transition but at twice the energy and of smaller amplitude (~100 pA). This corresponds to the second harmonic of the plasma transition, $2eV_J = 2h\nu_p$ (Figures 4b and 4d, blue dashed line), in which each Cooper pair tunneling through the spectrometer emits two photons of energy $h\nu_p$. This two-photon plasma transition $|-,0\rangle \to |-,2\rangle$ is represented by the blue dashed arrow in the energy ladder, Figure 4a. It is also possible to simultaneously excite the Andreev transition and the plasma mode (Figure 4a, purple dashed arrow). This type of transition $|-,0\rangle \to |+,1\rangle$, at $2eV_J = 2E_A + h\nu_p$, is observed in the spectra, Figures 4b and 4d, as a replica of the Andreev transition, shifted up by the plasma energy (purple dashed line). These transitions agree with the data everywhere except where two such two-photon processes coincide, near $\varphi = \pi$ and $2eV_J \approx 1.02\Delta$. There one observes a level repulsion (Figure 3a) or an avoided crossing (Figure 3b) depending on the relative position of the undressed states. In the spectra the region of instability obscures the hybridization effects at energy $2eV_J = h\nu_p$ but line traces slightly below confirm their existence (see Supplementary Information).

The experimental spectra are well described by a model based on the Andreev Hamiltonian[11] (see Section 2.3 of Supplementary Information). The eigenenergies of the SQUID Hamiltonian are determined by perturbation analysis and numerical diagonalisation.

The resulting transition energies are shown as black lines in Figure 4c and 4e. Only crossings of transition lines involving the same number of photons show significant hybridisation, in good agreement with the data. The rich structure predicted in the top part of the spectrum is not visible in the experiment because of the shunting by the SQUID capacitor. A quantitative description of the intensity and width of the transitions would require taking into account the coupling to the detector and the sources of dissipation.

Our results show that in addition to the phase difference, each conduction channel of a Josephson weak link possesses an internal degree of freedom similar to a spin-half. This Andreev pseudo-spin is unique as a microscopic degree of freedom intrinsically coupled to a superconducting circuit and whose energy is tunable over a wide range. Theoretical proposals for an Andreev qubit are based either directly on this pseudo-spin[11] or on the actual spin of quasiparticles trapped in the Andreev levels[12,13,16]. Their implementation requires reducing external sources of decoherence, something that could be achieved, in the circuit QED approach, by integrating a superconducting atomic contact in a high quality resonator[28,29]. Finally, in hybrid systems where spin-orbit and Zeeman interactions are also present, Andreev levels give rise to Majorana states whose detection is currently the subject of intense study[30].

## Methods

The sample is mounted in a bending mechanism (see Supplementary Figure 2d) anchored to the mixing chamber of a dilution refrigerator at 30 mK and housed inside a superconducting shield to reduce magnetic interference. Two SMA launchers connect it to the biasing and measuring lines which are heavily filtered. An electrically shielded small superconducting coil located directly above the sample is used to apply magnetic flux. A pusher actuated by a room temperature dc motor bends the sample and modifies the atomic contact configuration. The atomic contacts, tunnel junctions, and on-chip filters (alumina dielectric) are fabricated by electron-beam lithography and evaporation. Tunnel junctions are formed by double-angle evaporation and oxidation and have a bare plasma frequency of 14 GHz. Measurements of the SQUID and spectrometer current-voltage characteristics are made at low frequency (10−100 Hz) with room-temperature amplification. When the differential conductance of the spectrometer is smaller than $-1/R_b$, such as on the negative-slope side of the first plasma peak, biasing is unstable. This results in the absence of data in the grey regions above the plasma transition in Figures 2 and 3. At low voltages, there is another instability due to retrapping to the zero-voltage state. The peaks in the IVs which do not depend on the flux are subtracted from the measured spectra in the region $V_J>50\,\mu V$ as described in section 2.2 of Supplementary information. The theoretical spectra of Figure 4c and 4e are obtained by numerical diagonalisation of the Hamiltonian describing both the Andreev states and the plasma mode, which are coupled because they share the phase across the SQUID Josephson junction(see Supplementary information, section 2.3).

## Acknowledgments

We acknowledge technical assistance from P. Sénat and P.-F. Orfila, theoretical input from M. Houzet, help in the experiments by L. Tosi, and discussions with V. Shumeiko, A. Levy-Yeyati and within the Quantronics group. This work was supported by ANR contracts DOCFLUC and MASH, and by C'Nano. The research leading to these results has received funding from the People Programme (Marie Curie Actions) of the European Union's Seventh

Framework Programme (FP7/2007-2013) under REA grant agreement n° PIIF-GA-2011-298415.# References

[1] B.D. Josephson, Possible new effects in superconductive tunnelling. *Phys. Lett.* **1**, 251 (1962).

[2] S. Busch *et al.*, Measurements of T1-relaxation in ex vivo prostate tissue at 132 microT. *Magnetic Resonance in Medicine* **67**, 1138 (2012).

[3] Erik Lucero *et al.*, High-Fidelity Gates in a Single Josephson Qubit. *Nature Physics* **8**, 719–723 (2012).

[4] G. Wendin and V. S Shumeiko, Quantum bits with Josephson junctions. *Low Temp. Phys.* **33**, 724 (2007).

[5] J. M. Fink *et al.*, Climbing the Jaynes–Cummings ladder and observing its nonlinearity in a cavity QED system. *Nature* **454**, 315 (2008).

[6] Max Hofheinz *et al.*, Synthesizing arbitrary quantum states in a superconducting resonator. *Nature* **459**, 546 (2009).

[7] I. O. Kulik, Macroscopic quantization and proximity effect in S-N-S junctions. *Sov. Phys. JETP* **30**, 944 (1970).

[8] A. Furusaki and M. Tsukada, Dc Josephson effect and Andreev reflection. *Solid State Commun.* **78**, 299 (1991).

[9] C.W.J. Beenakker and H. van Houten, Josephson current through a superconducting quantum point contact shorter than the coherence length. *Phys. Rev. Lett.* **66**, 3056 (1991).

[10] Philip F. Bagwell, Suppression of the Josephson current through a narrow, mesoscopic, semiconductor channel by a single impurity. *Phys. Rev. B* **46**, 12573 (1992).

[11] A. Zazunov, V. S. Shumeiko, E. N. Bratus', J. Lantz, and G. Wendin, Andreev Level Qubit. *Phys. Rev. Lett.* **90**, 087003 (2003).

[12] N. M. Chtchelkatchev and Yu. V. Nazarov, Andreev Quantum Dots for Spin Manipulation. *Phys. Rev. Lett.* **90**, 226806 (2003).
10

# Supplementary information

is available on our website: http://iramis.cea.fr/drecam/spec/Pres/Quantro/static/

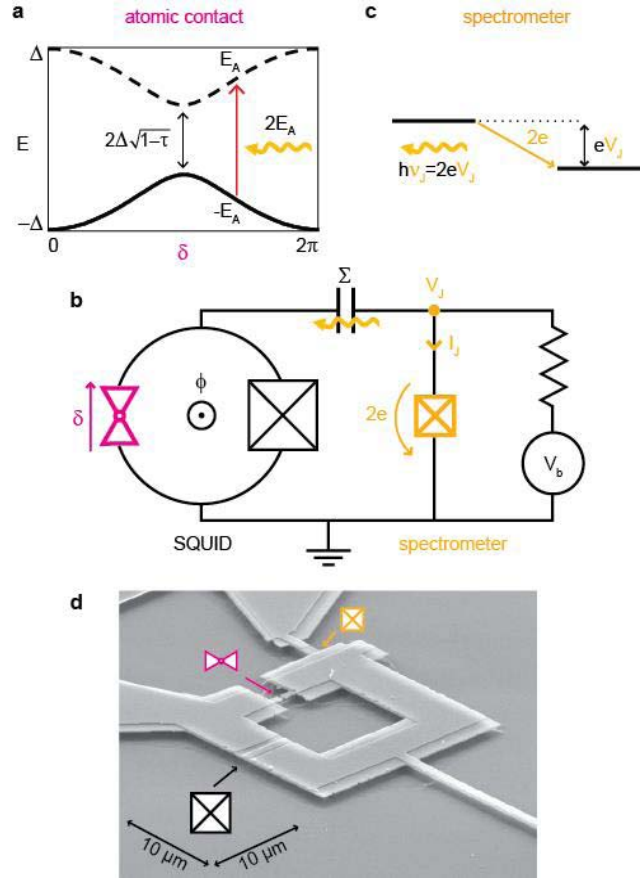

Figure **1. Principle of spectroscopy of the Andreev transition. a,** Phase (δ) dependence of the Andreev levels with energies $\pm E_A$ in a short transport channel of transmission $\tau$. Energy $\Delta$ is the superconducting gap. **b,** Simplified schematic of the setup. A voltage-biased Josephson junction (yellow checked box, critical current 48 nA) is used as a spectrometer: it acts both as a microwave source and a detector. The ac Josephson current (at frequency $\nu_J = 2eV_J/h$ set by the voltage $V_J$ across the junction), is coupled through capacitor $\Sigma$ to a SQUID formed by an atomic point contact (magenta triangles) and an ancillary Josephson junction (critical current 1 μA, 20 times larger than the typical critical current of a one-atom aluminum contact). Magnetic flux $\phi$ threading the loop imposes a phase $\delta \approx \varphi = 2\pi\phi/\phi_0$ across contact and determines the Andreev transition frequency of panel a. **c,** The absorption of a photon at this frequency is accompanied by the transfer of a Cooper pair through the spectrometer. **d,** Micrograph of the sample (at an angle of 45°) with spectrometer, suspended bridge to obtain the atomic contacts and SQUID Josephson junction.

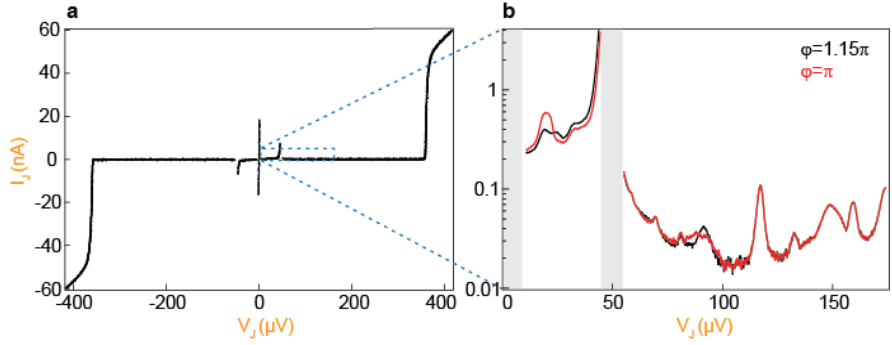

Figure 2. $I_J(V_J)$ **characteristic of the spectrometer coupled to SQUID with atomic contact AC2. a,** Large scale. **b,** Zoom of the sub-gap current for two values of the reduced flux $\varphi$. The grey regions at $V_J \leq 9\,\mu V$ and on the right-hand side of the peak at $V_J \approx 50\,\mu V$ are not accessible because the biasing is unstable. Parts of the spectra that change with $\varphi$ reveal energy absorption by the SQUID.

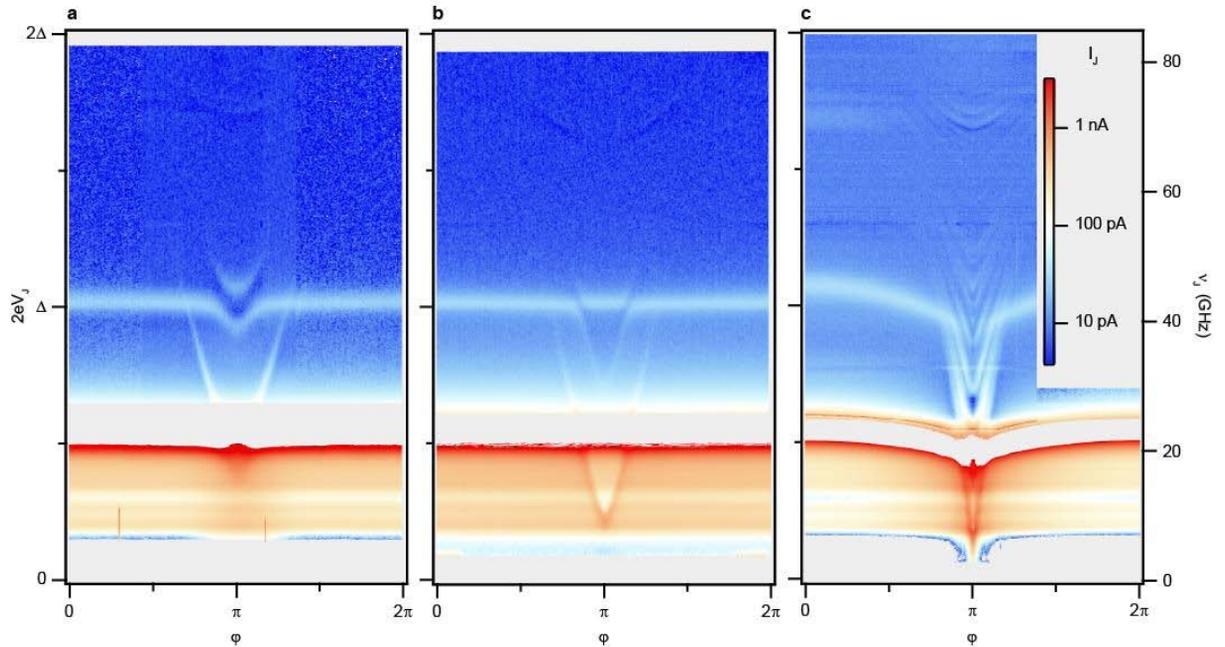

Figure 3. **Absorption spectra for 3 atomic contacts**: **a,** AC1 (transmissions 0.942, 0.26,...); **b,** AC2 (transmissions 0.985, 0.37,...); **c,** AC3 (more than 20 channels). The color encodes the current $I_J$ through the spectrometer, as a function of the reduced flux $\varphi$ and of the bias voltage $V_J$. The right axis gives the spectrometer frequency $\nu_J$ associated with $V_J$.

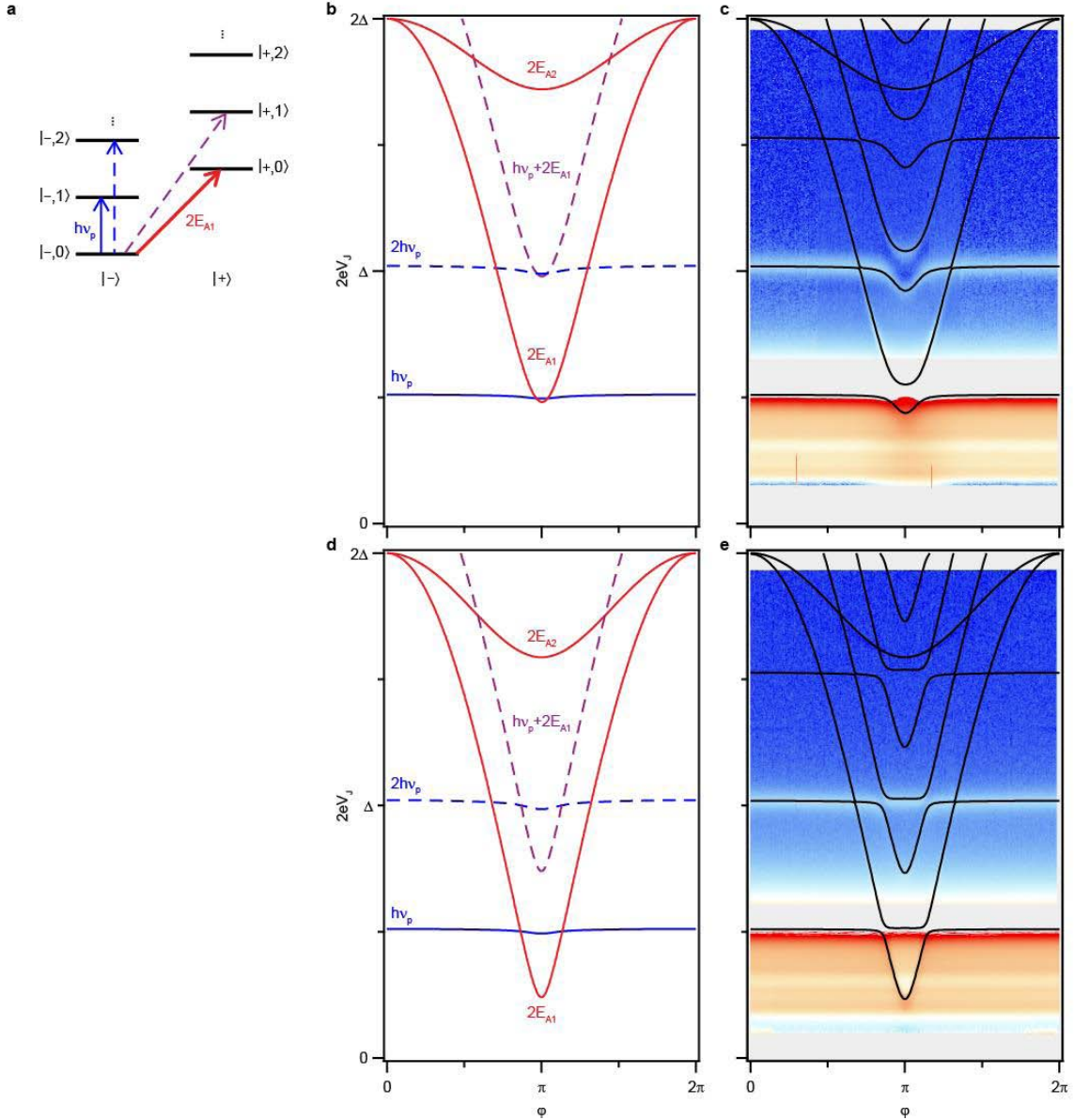

Figure **4. Interpretation of the absorption spectra. a,** Energy spectrum schematics for a single channel: each state is labeled $|-,n\rangle$ or $|+,n\rangle$ for the Andreev pair in the ground (−) or excited (+) state and $n$ photons in the plasma mode. **b,d,** predicted transitions for contacts AC1 and AC2. Red lines: transition energies $2E_A$ predicted from the channel transmissions (Eq. 1); blue lines: excitation energy for the plasma mode $h\nu_p$; blue dashed line: 2-photon plasma mode process $2h\nu_p$; purple dashed line: 2-photon process $2E_A+h\nu_p$ exciting both the Andreev and plasma transitions. **c,e,** Transition lines for contacts AC1 and AC2 calculated by diagonalisation of the full Hamiltonian, superimposed on the data.